\input{aipcheck}

\documentclass[
    ,final            
  ]
  {aipproc}

\layoutstyle{6x9}
\usepackage{graphicx}

\begin{document}

\title{A Census of White Dwarfs Within 40 Parsecs of the Sun}


\author{M.-M. Limoges\footnote{Visiting Astronomer, Kitt Peak National Observatory, 
National Optical Astronomy Observatory, which is operated by the Association of Universities 
for Research in Astronomy (AURA) under cooperative agreement with the National Science Foundation.}~}{ 
  address={D\'epartement de Physique, Universit\'e de Montr\'eal, C.P.~6128, 
Succ.~Centre-Ville, Montr\'eal, Qu\'ebec H3C 3J7, Canada}
}

\author{P. Bergeron}{
  address={D\'epartement de Physique, Universit\'e de Montr\'eal, C.P.~6128, 
Succ.~Centre-Ville, Montr\'eal, Qu\'ebec H3C 3J7, Canada},
}

\author{S\'ebastien L\'epine}{
  address={Department of Astrophysics, Division of Physical Sciences, American Museum of Natural History, 
Central Park West at 79th Street, New York, NY 10024},
}

\begin{abstract}
\noindent
Our aim is to compile a catalog of white dwarfs within 40 parsecs of the Sun, 
in which newly discovered objects 
would significantly increase the completeness of the current census. 
White dwarf candidates are identified from the SUPERBLINK proper motion database 
(\cite{lspm05}), which allows us to investigate stars down to a proper motion limit 
as low as 40 $mas~yr^{-1}$. The selection criteria and distance estimates are based on 
a combination of color-magnitude and reduced proper motion diagrams. Candidates with 
distances less than 50 parsecs are selected for spectroscopic follow-up. We present our 
preliminary sample of spectroscopically confirmed white dwarfs, as well as their atmospheric 
parameters. These parameters are obtained using the spectroscopic technique developed in 
\cite{bergeron92} for DA stars. DB, DQ, and DZ stars are also analyzed spectroscopically. 
For featureless spectra as well as those showing only 
$\rm{H}\alpha$, we perform a detailed photometric analysis of their energy distribution. 
 \end{abstract}

\maketitle


\section{Introduction}
\noindent
The current census of nearby white dwarfs is complete to within 20 parsecs from the Sun, and contains 
126 white dwarfs \citep{holberg08}. Since it represents a sample too small for detailed 
statistical analysis, there is a need to extend the complete sample of white dwarfs to a 
larger volume. 
Nearby white dwarfs have been traditionally found in catalogs of stars with 
large proper motions. In order to improve the statistics of the local white dwarf population, 
we have been hunting for white dwarfs in the SUPERBLINK catalog. \\

\noindent
The SUPERBLINK proper motion database \cite{lspm05} is based on a re-analysis of the POSS-I 
and POSS-II plates of the Digitized Sky Survey (20-45 yr baseline). Its 1.3 million stars in 
the Northern sky were identified with a detection level > $95\%$ down to $V=19$, and a proper 
motion limit as low as 0.04$'' yr^{-1}$. Because of its low proper motion limit, the SUPERBLINK 
sample effectively eliminates the kinematic bias of the Luyten catalogs, while detecting all 
white dwarfs down to the luminosity function turn-off.

\begin{figure}
  {\includegraphics[width=26pc]{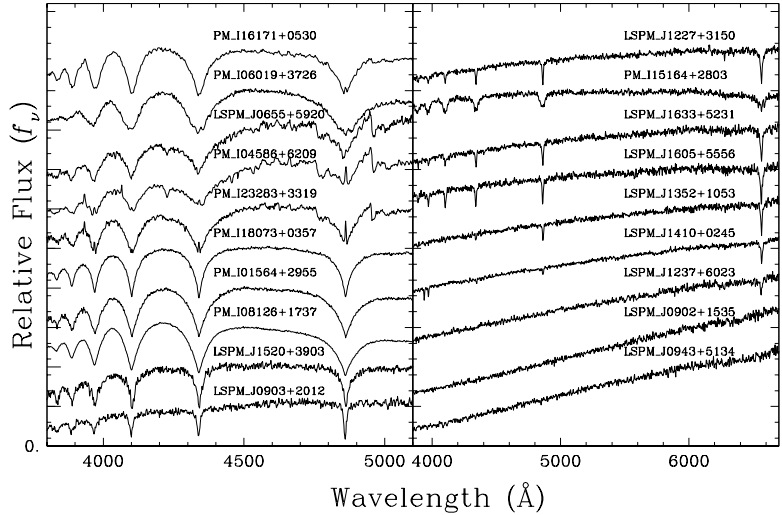}}
  \caption{Optical spectra for a subset of 19 stars from our new WD sample. 
 This sample shows 17 DA stars, and 2 featureless DC white dwarfs. The DA 
sample includes 4 DA+dM (PM I16171+0530, LSPM J0655+5920, PM I04586+6209 and 
PM I23283+3319) and two magnetic white dwarfs 
(PM I06019+3726 and PM I15164+2803).}
\end{figure}

\begin{figure}
  {\includegraphics[width=23pc]{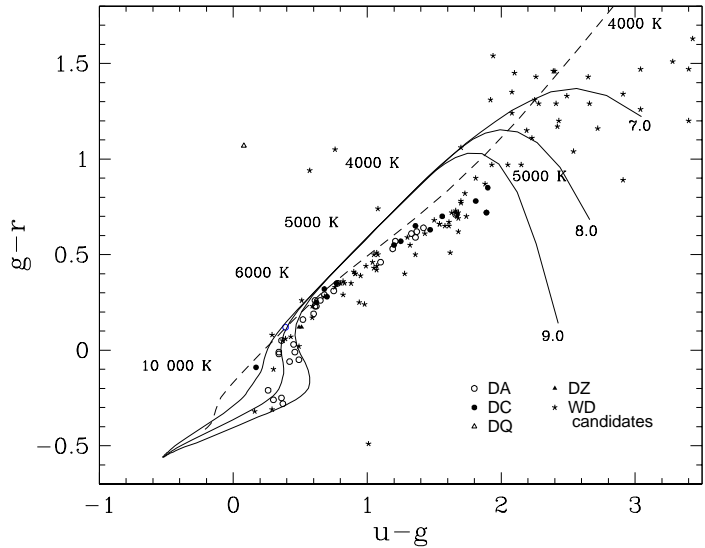}}
\caption{Color-color diagram showing the WD candidates and the spectroscopically confirmed WD from our study. 
The solid curves show model atmospheres for WD with a pure H atmosphere for log g=7.0, 8.0 and 9.0. 
The dashed curve represents a model for a pure He atmosphere with log g=8.0.}
\end{figure}

\section{Selection method}
\noindent
The first step 
of our analysis was to gather information about SUBERBLINK stars, as a complement to 
 photographic $B_F$, $R_I$ and $I_N$ measurements, J2000 coordinates, 
and proper motions available for each SUPERBLINK star (see \cite{lspm05}). 
Catalogs were queried in order to obtain GALEX far-UV ($F_{UV}$) and near-UV ($N_{UV}$) 
\citep{gil09}, SDSS $ugriz$ 
\citep{adel09} or 2MASS $JHK$ \citep{skrut06} photometry. Apparent $V$ magnitudes are 
estimated using the empirical relation $V=0.54B_F+0.45R_I$. Stars are then placed in 
$H_g$ vs $(g-z)$, $H_V$ vs $(V-J)$ or $H_V$ vs $(N_{UV}-V)$ reduced proper motion 
diagrams, depending on the available photometry for each object. When possible, 
cross-correlations between the diagrams are performed. Selection criteria are applied based on each star's 
position in the diagram (more details will be given in
 Limoges et al. 2011, in preparation). Finally, confirmed white dwarfs from \cite{ms06} 
are used to obtain color-magnitude calibrations, 
from which we determine an absolute magnitude and photometric distance for each candidate.
 Candidates with distance estimates less than 50 parsecs were selected for spectroscopic follow-up.\\

\begin{table}
\caption{Spectral types of our new white dwarfs sample}
\vspace{1pc}
\begin{tabular}{lcccccccr}
\hline
\hline
\fontsize{8}{10}
DA &  DB &  DC &  DQ &  DZ & DA+dM & DAmag  &  DA+DC &  Tot\\
\hline
76 &  0 &  43 &  2 &  4 &  8 &  4 &  2  &  137\\
\hline
\end{tabular}
\end{table}

\begin{table}
\caption{New White Dwarfs (Sample)}
\vspace{1pc}
\fontsize{8}{10}
\begin{tabular}{lccrrc}
\hline
\hline
Catalog Name&RA&DEC& $\mu$RA&$\mu$DEC&ST\\
 &(deg)&(deg)&($''$ yr$^{-1}$)&($''$ yr$^{-1}$) &\\
\hline
 LSPM J0021$+$2640 & 5.447206& 26.676771&$-$0.074&$-$0.322&DC   \\
 LSPM J0027$+$0542 & 6.902645&  5.700896& 0.266&$-$0.273&DA   \\
 PM I00331$+$4742& 8.293821& 47.703468&$-$0.060& 0.052&DA   \\
 LSPM J0055$+$5948 & 13.992868& 59.800690& 0.454&$-$0.061&DC   \\
 PM I01216$+$3440& 20.407501& 34.678669&$-$0.131& 0.017&DZ   \\
 LSPM J0127$+$7328 & 21.954573& 73.479942&$-$0.156& 0.075&DA   \\
 LSPM J0145$+$2918 & 26.435930& 29.306622& 0.529& 0.008&DC   \\
 LSPM J0148$+$3615 & 27.168560& 36.258621& 0.000&$-$0.231&DA   \\
 PM I01564$+$2955& 29.124598& 29.926645& 0.054&$-$0.089&DA   \\
 LSPM J0206$+$1836 & 31.561138& 18.606722& 0.784& 0.093&DC   \\
\hline
\end{tabular}
\end{table}

\noindent
All spectra have been obtained at Kitt Peak with the Steward Observatory 2.3-m telescope, 
and the NOAO Mayall 4-m and 2.1-m telescopes. 
The adopted configurations allow a spectral coverage 
of $\lambda\lambda$3200--5300 and $\lambda\lambda$3800--6700, 
at a mean resolution of $\sim 6$~\AA\ FWHM. A subset of DA and DC stars from our sample 
of new white dwarfs is shown in Figure 1. In Figure 2, a color-color diagram displays 
the confirmed white dwarfs and candidates for which $ugriz$ data were available. 
A summary of the spectral types for the 137 new white dwarfs is shown in Table 1. 
Finally, a sample of the detailed list of our new white dwarfs, 
including astrometry and spectral type is given in Table 2.\footnote{The complete version 
is available on www.astro.umontreal.ca/$\sim$limoges/ } 

\section{Results}
\begin{figure}
  \includegraphics[width=26pc]{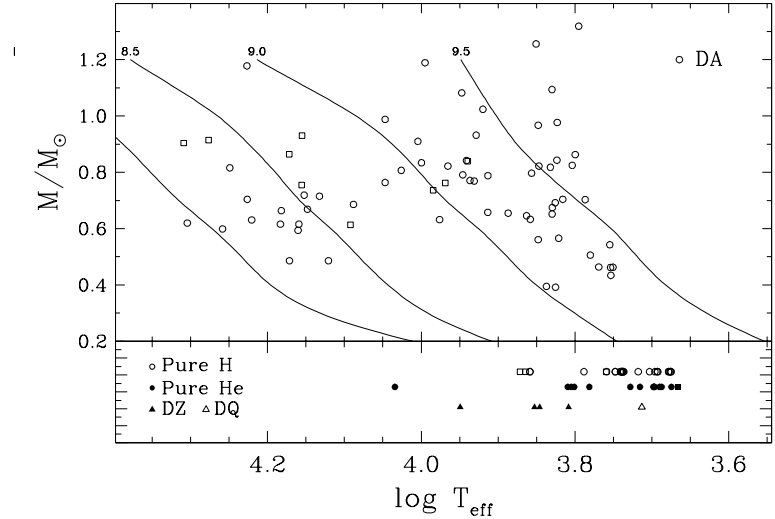}
  \caption{Top panel: Spectroscopic masses of DA white dwarfs in our sample as a function of $T_{\rm{eff}}$. 
Bottom panel: Photometric estimates of Teff; since no parallax measurements are available, 
$\log g$ is fixed at 8.0. The squares 
represent white dwarfs that were already known.}
  \label{correltm}
\end{figure}

\noindent
From our list of candidates, 137 white dwarfs\footnote{The DC companions in DA+DC systems
 are still candidates. They are not included in the total.} were found. 
Most of these objets were already 
part of the literature, without spectral type or white dwarf status confirmation. 
The atmospheric parameters ($T_{\rm eff}$ and $\log g$) of 114 stars were 
measured. DA stars were analyzed using the spectroscopic fitting technique 
described at length in \cite{lbh05} and references therein. Detailed photometric analyses of 
their energy distributions \citep{bergeron01} were performed for cooler objects showing only 
the ${\rm H}\alpha$ hydrogen line and featureless DC stars. Our results are summarized 
in Figure \ref{correltm} in a mass vs $\log T_{\rm eff}$ diagram.
Distances were then derived from the atmospheric parameters. It was found that 
82 stars are closer than 50 parsecs, and 57 are within 40 parsecs. 
The remaining 23 cool stars will be analyzed as soon as we acquire the photometric measurements. 
Figures 2 and \ref{correltm} demonstrate the effectiveness of our method in detecting 
cool white dwarfs. We are now reaching the bottom of the
luminosity distribution, and on our way to obtaining a statistically complete sample of nearby white
dwarfs.

\begin{theacknowledgments}
\noindent
We would like to thank the director and staff of Steward Observatory
for the use of their facilities. This work was supported in part by
the NSERC Canada and by the Fund FQRNT (Qu\'ebec). P.B. is a Cottrell
Scholar of the Research Corporation for Science Advancement.
\end{theacknowledgments}

\bibliographystyle{aipproc}   

\begin{thebibliography}{9}

\bibitem[Holberg et al.(2008)]{holberg08} Holberg, J.~B., Sion, E.~M., Oswalt, T., McCook, G.~P., Foran, S., \& Subasavage, J.~P. 2008, \emph{ApJ}, 135, 1225
\bibitem[L\'epine et al.(2005)]{lspm05} L\'epine, S., \& Shara, M.~M. 2005, \emph{AJ}, 129, 1483
\bibitem[Adelman-McCarthy et al.(2009)]{adel09}Adelman-McCarthy,J.K.; et al. 2009, VizieR On-line Data Catalog, 2294  
\bibitem[Gil de Paz et al.(2009)]{gil09} Gil de Paz et al. 2009, VizieR On-line Data Catalog, 21730185 
\bibitem[McCook $\&$ Sion (2006)]{ms06} McCook, G.P., $\&$ Sion, E.M. 2006 Vizier Online Data Catalog, 3235
\bibitem[Skrutskie et al.(2006)]{skrut06} Skrutskie, M.F. et al. 2006, \emph{AJ}, 131, 1163 
\bibitem[Liebert et al.(2005)]{lbh05} Liebert, J., Bergeron, P. \& Holberg, J. B. 2005 {\it ApJ}, {156} 47 
\bibitem[Bergeron et al.(1992)]{bergeron92} Bergeron, P., Saffer, R. \& Liebert, J. 1992, {\it ApJ}, {394} 247
\bibitem[Bergeron et al.(2001)]{bergeron01} Bergeron, P., Leggett, S. K., \& Ruiz, M. T. 2001, \emph{ApJS}, 133, 413

\end{thebibliography}

\end{document}